\documentstyle[12pt]{article}
\begin{document}
%%%%%%%%%%%%%%%%%%%%%%%%%%%%%%%%%%%%%%%%%%%%%%%%%%%
\def\thefootnote{\fnsymbol{footnote}}
\begin{flushright}
KANAZAWA-96-05  \\ 
March, 1996
\end{flushright}
%\vspace{ .7cm}
\vspace*{2cm}
\begin{center}
{\LARGE\bf Proton Stability and Small Neutrino Mass in String Inspired 
$E_6$ Models}\\
\vspace{1 cm}
{\Large  Daijiro Suematsu}
\footnote[3]{e-mail:suematsu@hep.s.kanazawa-u.ac.jp}
\vspace {1cm}\\
{\it Department of Physics, Kanazawa University,\\
        Kanazawa 920-11, Japan} \\   

{\it and}\\

{\it Theory Division, CERN CH-1211 Geneve 23, Switzerland}
\end{center}
\vspace{1.5cm}
{\Large\bf Abstract}\\  
%%%%%%%%%%%%%%%%%Abstract%%%%%%%%%%%%%%%%%%%%%%%%%%%
We propose a new possibility to realize simultaneously the sufficient 
proton stability and the interesting structure of neutrino mass matrix  
in superstring inspired $E_6$ models.
In this model the leptons and Higgs fields are assigned to a 
fundamental representation
{\bf 27} in the different way among generations. 
Two pairs of Higgs doublets naturally remain light from three generation ones
by imposing certain discrete symmetries, although all extra color triplets 
become sufficiently heavy. Under these symmetries suitable $\mu$-terms
to bring appropriate vacuum expectation values are prepared and the 
dangerous FCNC is avoidable. Some related phenomena to this model,
especially, the structure of neutrino mass matrix are also discussed.
\newpage
\setcounter{footnote}{0}
\def\thefootnote{\arabic{footnote}}
%%%%%%%%%%%%%%%%%Text%%%%%%%%%%%%%%%%%%%%%%%%%%%%%%%
\section{Introduction}
The unification of interactions is a very fascinating idea.
Although it brings many remarkable successes in
supersymmetric models\cite{n,ekn}, it also causes some difficulties because of
the strong constraints due to its unified group structure.
The existence of extra light color triplets generally causes serious problem
to the unified models since it makes proton unstable\cite{n}.
This problem occurs because extra color triplets are contained in the same 
multiplets together with the light doublet Higgs fields.
How to resolve this difficulty is one of the almost common issues of the 
realistic unified model buildings\cite{n,mis}.

It is well known that the same problem often annoys superstring inspired 
models too, although there are not necessarily the above mentioned multiplet 
structure.
In the realistic model buildings the existence of discrete symmetries and/or 
intermediate scales is often assumed in order to decouple these dangerous 
color triplet fields from ordinary quarks and leptons in the low energy 
world.

In superstring inspired $E_6$ models the same 
problem occurs. The full contents of ${\bf 27}$ of $E_6$ remain
massless in the low energy effective theory, although their multiplet 
structure is lost by the symmetry breaking due to the existence of the 
background fields on the extra dimensions\cite{w}.
A fundamental representation {\bf 27} contains extra color 
${\bf 3}$ and ${\bf 3}^*$ fields and 
the above mentioned triplet-doublet splitting problem appears.
Usually it is assumed the existence of an intermediate scale to make 
these extra color triplets heavy enough providing suitable discrete 
symmetries\cite{dks}. However, if we adopt such schemes it becomes difficult 
to give favorably small masses to neutrinos\cite{ms,sy2}.\footnote{
There are other type of models with no intermediate scale\cite{ceek}.
In such models extra color triplets are kept light but proton decay
is forbidden by discrete symmetries. Small neutrino mass is induced by the
loop effects\cite{ceek,rad}}
In string inspired $E_6$ models with an intermediate scale 
the proton stability and small neutrino mass seem not to be so 
easily reconciled.

In this paper in the certain type of string inspired $E_6$ models 
we propose a new possibility that small neutrino masses are
successfully introduced, although the proton stablity is guaranteed by 
making extra color triplet fields heavy.
The essential point of this scenario is the use of the freedom of the
field assignment in ${\bf 27}$ of $E_6$.
We will show how our scenario works in a very simple example and also 
discuss its phenomenological features briefly.

A fundamental representation ${\bf 27}$ of $E_6$ contains one generation 
quarks, leptons  and two singlets (${\bf 16} +{\bf 1}$ of $SO(10)$), 
a pair of Higgs doublets and a conjugate pair of extra color triplets 
(${\bf 10}$ of $SO(10)$) as shown in Table 1.
There is usually the extended gauge structure other than the standard model 
gauge group $G_{\rm SM}=SU(3)\times SU(2)\times U(1)$.
In the following discussion we consider the models whose gauge group
is $G_{\rm SM}\times U(1)^2$.
Two $G_{\rm SM}$ singlets have these extra $U(1)$ charges.
The gauge invariant superpotential for ${\bf 27}$ chiral superfields 
can be written by 
using the field notation presented in Table 1 as
\begin{eqnarray}
W&=&\lambda_1^{ijk} A_iA_jE_k+\lambda_2^{ijk} A_iB_jF_k
+\lambda_3^{ijk} A_iC_jG_k+\lambda_4^{ijk} A_iD_jH_k  \nonumber \\
&+&\lambda_5^{ijk} B_iC_jD_k+\lambda_6^{ijk} B_iE_jI_k 
+\lambda_7^{ijk} C_iE_jJ_k +\lambda_8^{ijk} D_iE_jK_k \nonumber \\
&+&\lambda_9^{ijk} F_iG_jK_k+\lambda_{10}^{ijk}F_iH_jJ_k 
+\lambda_{11}^{ijk}G_iH_jI_k +\cdots ,
\end{eqnarray}
where indices $i,j$ and $k$ represent the generation.
The ellipses stand for nonrenormalizable terms.
As is seen from Table~1, each pair of $(C, D), (G, H)$ and $(J, K)$ has the
same quantum numbers of $G_{\rm SM}$, respectively. 
This means that there remains the freedom how to assign the physical
fields to them if suitable phenomenological conditions are satisfied.
Usually the assignment is adopted so as to guarantee the existence of
the following necessary terms in W:
\begin{equation}
Q_i\bar U_j H^2,\qquad Q_i\bar D_jH^1,\qquad L_i\bar E_jH^1,
\qquad SH^1H^2.
\end{equation}
The first three terms are Yukawa couplings which induce quarks 
and charged leptons mass. The last one brings so-called $\mu$-term
after a singlet $S$ gets a suitable vacuum expectation value(VEV).
In the conventional assignment\cite{dks,ms}, moreover, the same assignment 
is assumed to be applyed to all three generations.
However, in principle, there is no necessity for such a field assignment.
We can adopt different ones for each generation.
In fact, such an unconventional assignment has been proposed in 
ref.\cite{nar} within no intermediate scale models. 
As pointed out in it, there appear some novel phenomena associated with such
assignments, for example, extra color triplets remain light, 
the neutrino mass appears at one loop
level, the extra $U(1)$ interaction loses its universality among generations 
and so on. However, if suitable discrete symmetries are imposed and 
parameters are also appropriately chosen, then they can be consistent with 
all experimental constraints at the present stage.

There are other typical models with the extended abelian gauge
structure and an intermediate scale in the
string inspired $E_6$ models. In this kind of models the extra fields
generally have very different features in comparison with ones in 
\cite{nar} because of the existence of intermediate scale. 
It is interesting and also useful to study what effects are induced 
by such unconventional assignments in the models with an intermediate
scale from the viewpoint of the model building.
In the followings, we will show that in the intermediate scale models
it is possible to prohibit the fast proton decay due to the
triplet-doublet splitting and also give neutrinos 
small masses in the very simple way.
It should be noted that in our model these features are largely dependent 
on the introduction of the intermediate scale, which is the crucial 
difference from the model in ref.\cite{nar}.\footnote{
There are some arguments that the existence of the intermediate scale is 
phenomenologically unfavorable\cite{ceek}. In our model, however, 
there are no such problems. The excessive entropy production associated
with flat directions is the common problem in superstring models\cite{int}.}
In particular, the introduction of an intermediate scale makes it
possible to present the neurino mass matrix with the large Majorana mass 
of a right handed neutrino.
In this neutrino mass matrix the seesaw mechanism\cite{see} works
and then the neutrino sector can realize the wide range mass scales 
by the collaboration with the loop effects. 
These mass scales may simultaneously solve the neutrino 
problems\cite{solar,atm,dark}, which have recently attracted much
interests of many authors. This feature cannot be seen in the model of 
ref.\cite{nar} in which only loop effects induce the neutrino masses.  

\section{A model with an unconventional field assignment}
We consider a simple example of models which satisfy some properties which
we require.
This model is characterized by two features.
One is the unconventional field assignment and the other is the existence
of a massless singlet which associates with its conjugate field.
The latter makes it possible to introduce an intermediate scale as explained
in the next part.
The field assignment of our model is presented in Table 1. 
We take the same field assignment for the first 
two generations but the third generation is assigned in the different way.

As is well known in the Wilson loop breaking mechanism 
in superstring theories\cite{w},
there can generally exist massless conjugate pairs of chiral superfields 
$({\bf R}, {\bf\bar R})$ other than ${\bf 27}$\cite{dks,ms}. Here 
${\bf R}$ represents some components of ${\bf 27}$ given in Table 1.
Taking this fact into account, we assume the existence of a conjugate pair of
$G_{\rm SM}$ singlet chiral superfields $(J, \bar J)$ and represent them as
$({\cal J},\bar{\cal J})$.\footnote{
The systematic study of this kind of spectrum has been done in ref.\cite{ms}.
From its results it is found that two different types of singlets
(for example, $J$ and $K$) can not be massless with its conjugate fields 
simultaneously, at least in the case that the gauge structure is $G_{\rm SM}
\times U(1)^2$.
This is because Wilson loop can not be constructed to be orthogonal
to both $J$ and $K$.}

The superpotential W can be devided into a $({\cal J},\bar{\cal J})$
independent part $W_0$ and a dependent part $W_{\cal J}$.
They are written down explicitly by using the physical fields notation given
in Table~1,
\begin{equation}
W=W_0 + W_{\cal J},
\end{equation}
\begin{eqnarray}
W_0&=&\lambda_1^{ijk}Q_iQ_jg_k+\lambda_2^{ijk}Q_i\bar U_jH^2_k
+\lambda_3^{ij(\alpha 3)}Q_i\bar g_j
\left( \begin{array}{c}L_\alpha \\ H^1_3 \\
         \end{array}\right)
+\lambda_4^{ij(\alpha 3)}Q_i\bar D_j
\left( \begin{array}{c}H^1_\alpha \\ L_3 \\
         \end{array}\right) \nonumber \\
&&+\lambda_5^{ijk} 
\bar U_i\bar g_j\bar D_k
+\lambda_6^{ijk}\bar U_ig_j\bar E_k
+\lambda_7^{ij(\alpha 3)}\bar g_ig_j
\left( \begin{array}{c}S_\alpha \\ \bar N_3 \\
         \end{array}\right)
+\lambda_8^{ij(\alpha 3)}\bar D_ig_j
\left( \begin{array}{c}\bar N_\alpha \\ S_3 \\
         \end{array}\right) \nonumber \\
&&+\lambda_9^{i(\alpha 3)(\beta 3)}
H^2_i\left( \begin{array}{c} L_\alpha \\ H^1_3 \\
         \end{array}\right)
\left( \begin{array}{c}\bar N_\beta \\ S_3 \\
         \end{array}\right)
+\lambda_{10}^{i(\alpha 3)(\beta 3)}
H^2_i\left( \begin{array}{c}H^1_\alpha \\ L_3 \\
         \end{array}\right)
\left( \begin{array}{c}S_\beta \\ \bar N_3 \\
         \end{array}\right)  \nonumber \\
&&+\lambda_{11}^{(\alpha 3)(\beta 3)i}
\left( \begin{array}{c}L_\alpha \\ H^1_3 \\
         \end{array}\right)
\left( \begin{array}{c}H^1_\beta \\ L_3 \\
         \end{array}\right) \bar E_i+...,    \\
W_{\cal J}&=& \lambda_7^{ij{\cal J}}\bar g_ig_j{\cal J}
+\lambda_{10}^{i(\alpha 3){\cal J}}H_i^2
\left( \begin{array}{c} H^1_\alpha \\ L_3 \end{array}\right){\cal J}
+{\lambda_{12}^{(\alpha 3)\bar{\cal J}} \over M_{\rm pl}}
 \left[\left(\begin{array}{c} S_\alpha \\ \bar N_3\end{array}\right)
\bar {\cal J} \right]^2  \nonumber \\
&&+{\lambda_{13}^{{\cal J}\bar{\cal J}} \over M_{\rm pl}^{2n-3}}
 \left[{\cal J}\bar{\cal J}\right]^n +...,
\end{eqnarray} 
where $n$ corresponds to the dimension of the lowest order gauge invariant
allowed nonrenormalizable term which contains ${\cal J}$ and ${\bar{\cal J}}$.
The ellipses represent higher order nonrenormalizable terms.
The indices $\alpha$ and $\beta$ stand for the first and second
generations. 

The conjugate pair $({\cal J}, \bar{\cal J})$ has an opposite charge of the 
extra $U(1)$ and then, as is well known, there is a D-flat direction 
$|\langle{\cal J}\rangle |=|\langle\bar{\cal J}\rangle |$.
If the negative soft squared mass $-m_S^2$ for the scalar component of
${\cal J}$ is induced as a result of 
supersymmetry breaking and also the radiative effects due to the Yukawa
couplings $\lambda_7^{ij{\cal J}}$ and $\lambda_{10}^{i(\alpha 3){\cal J}}$, 
VEVs of ${\cal J}$ and $\bar{\cal J}$ will be produced through the 
$\lambda_{13}^{{\cal J}\bar{\cal J}}$ term in $W_{\cal J}$ as 
follows\cite{dks,ms},
\begin{equation}
|\langle{\cal J}\rangle |=|\langle \bar{\cal J}\rangle | \sim 
\left(\left(\lambda_{13}^{{\cal J}\bar{\cal J}}\right)^{-1}
M_{\rm pl}^{2n-3}m_S \right)^{1 \over 2n-2}.
\end{equation}
If Yukawa couplings $\lambda_7^{ij{\cal J}}$ and 
$\lambda_{10}^{i(\alpha 3){\cal J}}$
in $W_{\cal J}$ are suitably arranged, all extra color triplets 
$g_i$, $\bar g_i$ and only one pair of Higgs doublets 
$(H^1_1,H^2_1)$ become heavy due to these VEVs.\footnote{
As discussed later, in order to introduce the suitable neutrino mass structure
we need two pair of light Higgs doublets. The possibility of these arrangements
of Yukawa couplings will be justified by the later argument of discrete 
symmetries.}
For example, if $n=3$ and $\lambda_{13}^{{\cal J}\bar{\cal J}}=O(1)$, 
$|\langle{\cal J}\rangle |$ becomes large enough as $\sim 10^{15}$GeV. 
Thus the proton decay process mediated by $g_i$ and $\bar g_i$ can be 
sufficiently suppressed\cite{proton}. 
Moreover, through the $\lambda_{12}^{(\alpha 3){\cal J}}$ term 
$S_\alpha$ and $\bar N_3$ can get
the mass of order of 
$\lambda_{12}^{(\alpha 3){\cal J}}
{|\langle{\cal J}\rangle |^2 \over M_{\rm pl}}$.\footnote{
This possibility has already suggested in \cite{sy2}. However,
the conventional field assignment was used there so that the extra color 
triplets could not be heavy. There is a proposal\cite{cl} to introduce
two D-flat directions, each of which is responsible for the heavy mass
of extra color triplets and right handed neutrinos, respectively.
Such two D-flat directions, however, seem to be difficult to exist 
simultaneously within the framework of usual Wilson loop breaking
as mentioned in footnote~3.} 
As discussed later, $N_3$ plays a role of the heavy right handed neutrino.
For $S_\alpha$ we assume that only $S_1$ becomes super heavy.\footnote{
We will discuss the consistency of this assumption related to
the discrete symmetries later.}
This assumption is also related to the neutrino mass production.
It should be noted that these phenomena can simultaneously occur 
because of the unconventional field assignment adopted here.

Now we can write down the effective superpotential $W_{\rm light}$
of light fields,
\begin{eqnarray}
W_{\rm light}&=&W_1+W_2+W_3, \\
W_1&=&\lambda_2^{ij(23)}Q_i\bar U_j
\left( \begin{array}{c}H^2_2 \\ H^2_3 \\ \end{array}\right)
+\lambda_4^{ij2}Q_i\bar D_jH^1_2
+\lambda_{11}^{\alpha 2i }L_\alpha H_2^1\bar E_i
+\lambda_{11}^{33i}H_3^1L_3\bar E_i  \nonumber \\
&&+\lambda_9^{(23)33}
\left( \begin{array}{c}H^2_2 \\ H^2_3 \\
         \end{array}\right)H^1_3 S_3 
+\lambda_{10}^{(23)22}
\left( \begin{array}{c}H^2_2 \\ H^2_3 \\
         \end{array}\right) H^1_2S_2 
+\lambda_{10}^{(23)33}
\left( \begin{array}{c}H^2_2 \\ H^2_3 \\
         \end{array}\right)L_3\bar N_3,  \\
W_2&=&\lambda_4^{ij3}Q_i\bar D_jL_3 
+\lambda_{11}^{\alpha 3i}L_\alpha L_3\bar E_i
+\lambda_{11}^{32i}H^1_3H^1_2\bar E_i, \\
W_3&=&\lambda_9^{(23)\alpha\beta}
\left( \begin{array}{c} H^2_2 \\ H^2_3 \\
         \end{array}\right)L_\alpha \bar N_\beta
+\lambda_9^{(23)\alpha 3}
\left( \begin{array}{c} H^2_2 \\ H^2_3 \\
         \end{array}\right)L_\alpha S_3
+\lambda_9^{(23)3\alpha}
\left( \begin{array}{c} H^2_2 \\ H^2_3 \\
         \end{array}\right)H^1_3 \bar N_\alpha \nonumber \\
&&+\lambda_{10}^{(23)32}
\left( \begin{array}{c}H^2_2 \\ H^2_3 \\
         \end{array}\right)L_3S_2
+\lambda_{10}^{(23)23}
\left( \begin{array}{c} H^2_2 \\ H^2_3 \\
         \end{array}\right)H^1_2\bar N_3,
\end{eqnarray}
where we add the terms relevant to $\bar N_3$ although it is heavy.\footnote{
Although $S_1$ can play the same role as $\bar N_3$ at this stage,
we distinguish them here. This treatment will be also be justified
by the introduction of discrete symmetries.}
All necessary terms which induce
quark and charged lepton masses and $\mu$-terms are contained in $W_1$.
Typical $R$-parity violating terms are contained in $W_2$.
In $W_3$ phenomenologically dangerous terms are included. They should 
be forbidden by suitable discrete symmetries since they cause the 
unwanted masses and
mixings among neutral fermions after doublet Higgses get VEVs.
It is the most interesting problem what kind of neutrino masses
are induced in this $W_{\rm light}$.
It also gives the most important criterion to introduce the discrete 
symmetries for select terms from $W_{\rm light}$.

\section{Discrete symmetries and neutrino masses}
In this section we examine the discrete symmetries and their relation to
the neutrino masses and set up our model definitely.
In order to introduce the discrete symmetries we will impose the following 
conditions:\\
(i)~all necessary terms in $W_1$ are kept as invariant ones,\\
(ii)~to avoid the FCNC problem in the quark sector we require that only 
$H_2^2$ couples to the up-quark sector
\begin{equation}
\lambda_2^{ij3}=0,
\end{equation}
(iii)~$W_2$ which includes usual R-parity violating terms is forbidden
\begin{equation}
\lambda_4^{ij3}=\lambda_{11}^{\alpha 3i}=\lambda_{11}^{32i}=0,
\end{equation}
(iv)~all extra color triplets can become heavy through the couplings with
${\cal J}$ in $W_{\cal J}$
\begin{equation}
\lambda_7^{ij{\cal J}}\not=0,
\end{equation}
(v)~only $S_2$ in the singlets $(S_\alpha, \bar N_3)$ and two pairs of doublet 
Higgses $(H_2^1,H^2_2),~(H_3^1, H_3^2)$ remain massless so that they are 
forbidden to couple with ${\cal J}$ and $\bar{\cal J}$
\begin{equation}
\lambda_{10}^{2(23){\cal J}}=\lambda_{10}^{3(23){\cal J}}
=\lambda_{12}^{2\bar{\cal J}}=0.
\end{equation}
Under these conditions we will find out appropriate discrete symmetries
and study their results.
Before giving such an example, it will be useful to present 
some features of the
terms in $W_{\rm light}$ relevant to the neutrino masses. 

As seen from eq.(5), in the present model only the third generation neutrino
$N_3$ gets the large Majorana mass and then the seesaw mechanism\cite{see} 
works for Dirac masses related to $\bar N_3$. 
From this point of view, to avoid large 
left handed neutrino masses it is necessary to impose
\begin{equation}
\lambda_9^{(23)\alpha\beta}=\lambda_9^{(23)\alpha 3}=
\lambda_{10}^{(23)32}=0.
\end{equation}
Under these assumptions $L_3\bar N_3$ Dirac mass and $\bar N_3\bar N_3$ 
large Majorana mass alone exist at tree level.
However, due to the radiative corrections based on the remaining interactions 
of $W_{\rm light}$, $L_\alpha \bar N_3$ Dirac masses and 
$\bar N_\alpha \bar N_3$, 
$\bar N_\alpha \bar N_\beta$ Majorana masses are induced through the one 
loop diagram shown in Figs.1-3.
If we assume that soft supersymmetry breaking parameters take the universal
value $O(m_{3/2})$,
their magnitudes are roughly estimated as\cite{ceek,rad},
\begin{eqnarray}
&&m \sim {A \over 32\pi^2}\lambda_{11}^{\alpha 23}
\lambda_{10}^{(23)33}m_\tau \quad\qquad ({\rm for}~L_\alpha \bar N_3),
\nonumber \\
&&m \sim {A \over 32\pi^2}\lambda_9^{(23)3\alpha}
\lambda_{10}^{(23)23}m_{\tilde H} \qquad ({\rm for}~
\bar N_\alpha \bar N_3),
\nonumber \\
&&m \sim {A \over 32\pi^2}\lambda_9^{(23)3\alpha}
\lambda_9^{(23)3\beta}m_{\tilde H} 
\qquad ({\rm for}~\bar N_\alpha \bar N_\beta),
\end{eqnarray}
where $m_\tau$ and $m_{\tilde H}$ are the masses of tau and corresponding 
charginos. The soft breaking A-terms are parametrized as $Am_{3/2}$.
In the present model there are fruitful structures in the Higgs
sector. It should be noted that
as its result there may be the tree level contributions to these
masses which are not explicitly presented here. This can be easily
seen by replacing the
Higgs internal lines into their VEVs in Figs.1$\sim$3.
However, their relative largeness completely depends on the 
values of soft supersymmetry breaking parameters and Higgs VEVs.
At one-loop level $L_i\bar N_\alpha$ Dirac masses and $L_iL_j$ Majorana masses
are not induced under the present assumptions.
On the other hand, Yukawa couplings $\lambda_9^{(23)3\alpha}$ and 
$\lambda_{10}^{(23)23}$
induce the mixings between Higgsinos and right handed neutrinos.
These mixings can largely affect the $N_\alpha$ Majorana masses,
although their effects on the $N_3$ Majorana mass are negligible.
In order for such effects to be of order $10^{-1}$~eV,
the relevant Yukawa couplings $\lambda_9^{(23)3\alpha}$ should be
less than $O(10^{-4})$ if we take the Higgsino masses as $\sim 100$~GeV.

Now we look for the discrete symmetries which satisfy the conditions
(i) $\sim$ (v). These discrete symmetries should not be broken by the VEVs
of ${\cal J}$ and $\bar{\cal J}$ so that they must not have their charges.
Taking account of 
these, we can find a simple but interesting example of such discrete 
symmetries and $W_{\rm light}$ invariant under it.
Such an example is $Z_2\times Z_2\times Z_n~(n \ge 3)$ and the charge 
assignment for each field is\footnote{
We systematically searched such solutions as satisfying conditions (i) $\sim$ 
(v) within $Z_2 \times Z_2\times Z_n$ type discrete symmetries providing that
the quark sector transforms as simple as possible under them. 
The promising solution is very restricted
and the following one seems to be almost unique.
It seems to be difficult that the condition (15) is satisfied, simultaneously.
An unwanted term $\lambda_9^{3\alpha 3}H_3^2L_\alpha S_3$ can not be 
forbidden only by the present discrete symmetry. Although we may need to impose
more complicated discrete symmetry to prohibit it, we only assume here that 
this Yukawa coupling is accidentally zero.
If it happens, it can be checked that the dangerous mixings induced by its
existence are not caused by the one-loop effects.} 
\begin{eqnarray}
&&H_3^2(-1, 1, \alpha),\quad H_2^1(1, -1, 1), \quad 
H_3^1(1, 1, \alpha\beta^{-1}),\quad S_2(-1, -1, \alpha^{-1}),\nonumber \\
&&S_3(1, 1, \alpha^{-1}\beta), \quad
L_\alpha(-1, 1, \beta^{-1}), \quad L_3(-1, -1, \alpha^{-1}), \nonumber \\
&&\bar E_i(-1, -1, \beta), \quad
 \bar N_\alpha(-1, 1, \alpha^{-2}\beta), \quad \bar N_3(1, -1, 1), 
\quad \bar D_i(1, -1, 1), 
\end{eqnarray}
where we represent the charges of each field as $(p, q, r)$ where $p$, $q$
and $r$
are charges of $Z_2$'s and $Z_n$. Nontrivial 
$Z_n$ elements $\alpha$ and $\beta$ satisfy $\alpha^n=\beta^n=1$.
All other fields in Table~1 including ${\cal J}$ and $\bar{\cal J}$
are invariant under this discrete symmetry.
The superpotential $W_{\rm light}$ of light fields can be written as  
\begin{eqnarray}
W_{\rm light}&=&\lambda_2^{ij2}Q_i\bar U_jH^2_2 
+\lambda_4^{ij2}Q_i\bar D_jH^1_2
+\lambda_{11}^{\alpha 2i }L_\alpha H_2^1\bar E_i
+\lambda_{11}^{33i}L_3 H_3^1\bar E_i  \nonumber \\
&&+\lambda_9^{233}H^2_2H^1_3 S_3 
+\lambda_{10}^{322}H^2_3H^1_2S_2 
+\lambda_{10}^{333}H^2_3L_3\bar N_3,  \nonumber \\
&&+\lambda_9^{33\alpha}H^2_3H^1_3 \bar N_\alpha 
+\lambda_{10}^{223}H^2_2H^1_2\bar N_3.
\end{eqnarray}
It is noticable that this superpotential contains all necessary terms and
some nice features. Although there are two Higgs doublet pairs
$(H_2^1, H_2^2)$ and $(H_3^1, H_3^2)$ which can be expected to get
VEVs through the existence of $\mu$-term couplings $\lambda_9^{233}$ and
$\lambda_{10}^{322}$, up and down quarks couple a different Higgs field
respectively and then the large FCNC in the quark sector can be avoidable as
the minimal supersymmetric standard model(MSSM).
Although the structure of $W_{\rm light}$ is similar to the one of the MSSM,
the differences from the MSSM appear in lepton and Higgsino sectors.
The charged leptons have Yukawa couplings with two Higgs fields and
there are two types of $\mu$-terms. The detailed study of these aspects is 
beyond the scope of this paper but the further investigation of FCNC in the
charged lepton sector and also the phenomenology of Higgsino sector will
be necessary.

The additional terms in this $W_{\rm light}$ result in the interesting features
in neutrino masses as noted in the previous part. Here we discuss the 
neutrino mass matrix in this model in some details.  
At the stage of the first approximation
the neutrino mass matrix induced from this $W_{\rm light}$ can be written in
the $(L_\alpha, L_3, \bar N_\alpha, \bar N_3)$ basis as,
\begin{equation}
\left( \begin{array}{cccc}
 0&0&0&m_{\alpha 3}\\ 0&0&0&m_{33}\\ 
 0&0&M_{\alpha\beta}&M_{\alpha 3}\\ m_{\alpha 3}&m_{33}&M_{\alpha 3}&M_{33}\\
\end{array}\right),
 \end{equation} 
where using formulae (16) each element is expressed as,
\begin{eqnarray}
\displaystyle
&&m_{33}\sim\lambda_{10}^{333}\langle H_3^2\rangle, \nonumber \\
&&m_{\alpha 3}\sim {\rm max}\left[
\lambda_{11}^{\alpha 23}
\lambda_{10}^{333}{\langle H_2^1\rangle\langle H_3^2\rangle \over
m_\tau},~
{A \over 32\pi^2}\lambda_{11}^{\alpha 23}
\lambda_{10}^{333}m_\tau
\right], \nonumber \\
&&M_{\alpha 3}\sim {\rm max}\left[
 \lambda_9^{33\alpha}\lambda_{10}^{223}
{\langle H_2^2\rangle\langle H_3^2\rangle \over
m_{\tilde H}},~
{A\over 32\pi^2}\lambda_9^{33\alpha}
\lambda_{10}^{223}m_{\tilde H} \right], \nonumber \\
&&M_{\alpha\beta}\sim \lambda_9^{33\alpha}\lambda_9^{33\beta}
{\left[{\rm max}(\langle H_3^2\rangle , \langle H_3^1\rangle)\right]^2 
\over m_{\tilde H} }, \nonumber \\ 
&&M_{33}\sim \lambda_{12}^{3\bar{\cal J}}{|\langle {\cal J}\rangle |^2\over 
M_{\rm pl}}.
\end{eqnarray}
The tree level contributions are explicilty presented in $m_{\alpha 3}$ and
$M_{\alpha 3}$. 
Although there is no contribution to $M_{\alpha\beta}$ from Fig.3,
$M_{\alpha\beta}$ is induced as the result of the mixings with Higgsinos.
These elements should satisfy the condition 
\begin{equation}
M_{\alpha\beta} \ll m_{\alpha 3}, m_{33}, M_{\alpha 3} \ll M_{33}.
\end{equation} 
This type of matrix has three nonzero light mass eigenvalue other than 
$M_{33}$. 
Based on the analyses of solar neutrino\cite{solar}, 
atmospheric neutrino\cite{atm} and various 
cosmological observations\cite{dark}, it seems to be preferable to consider
that there are hierarchical three typical mass scales related to the 
light neutrino sector, that is, $\sim$10~eV (dark matter), 
$\sim 10^{-1}$eV (atmospheric neutrino) and $\sim 10^{-3}$eV (solar
neutrino) from the viewpoint of mass differences. 
If there is hierarchical structures in $M_{\alpha\beta}$,
these appropriate mass scales will be induced in the light neutrino 
sector through the collaboration with
the seesaw mechanism based on a right handed Majorana neutrino
mass $M_{33}$ as,
\begin{equation}
M_{11}(\sim 10^{-3}~{\rm eV}) \ll 
M_{22}(\sim 10^{-1}~{\rm ev}) \ll 
{m_{33}^2 \over M_{33}}(\sim 10~{\rm eV}) 
\end{equation}
In that case, for example,
two flavor oscillations such as $\nu_e \leftrightarrow \bar N_1$ 
and $\nu_\mu \leftrightarrow \bar N_2$
may solve the solar and atmospheric neutrino problems and also 
$\nu_\tau$ will be a candidate of the dark matter.
For such an identification, the elements of eq.(19) should take the values
\begin{equation}
m_{33}\sim 1~{\rm GeV},\quad M_{23} ~{^<_\sim}~ 10^{-1}~{\rm GeV}, 
\quad M_{13}~{^<_\sim}~ 10^{-2}~{\rm GeV} 
\end{equation}
for $M\sim 10^8$~GeV.\footnote{
This rather small right handed neutrino mass can be realized by taking 
$\lambda_{12}^{3\bar{\cal J}}\sim 10^{-3}$ even 
if $\langle{\cal J}\rangle\sim 10^{15}$GeV 
which can guarantee the proton longevity. Small value of 
$\lambda_{12}^{3\bar{\cal J}}$
may be explained by the fact that in a certain type of string models
nonrenormalizable terms are induced by nonperterbative
effects\cite{cw}.}
The arguments here are concentrated on the mass scales in the neutrino 
sector and the mixing angles are not discussed. 
For the study of mixing angles
more complicated loop effects should be taken into account, which will 
fill the places of zero components in eq.(19).\footnote{
In this type of neutrino mass matrix, the neutrino oscillation
phenomena are studied in ref.\cite{s}.}
Such effects can be expected to be the same order as $M_{\alpha\beta}$
or smaller than it. Thus the qualitative feature of the above
arguments will not be changed.
This subject, however, is beyond the scope of the present paper and
we will not discuss it farther.

Now we examine the possibility to realize this mass hierarchy (21)$\sim$(23) 
in the neutrino mass matrix (19) more concretely.
Using formulae (20) to estimate the elements of this matrix, we 
take parameters in the following way.
For Yukawa couplings $\lambda_{10}^{223}$ and $\lambda_{10}^{333}$ 
there is no phenomenological constraints and then we can take 
it as $\sim O(1)$.
Here it is useful to note that $m_{33}$ depends on $\langle H_3^2\rangle$
which is irrelevant to the quarks and charged leptons masses and 
can be taken to be small enough as $\sim$ 1 GeV. 
The consistency to the charged lepton mass eigenvalues requires that
 $\lambda_{11}^{123}~{^<_\sim}~6\times 10^{-4}$ and 
$\lambda_{11}^{223}~{^<_\sim}10^{-2}$ under the assumption of $\langle 
H^1_2\rangle \sim 50$~GeV. 
As mentioned before, $M_{\alpha\beta}$ comes from the mixing 
between $\bar N_\alpha$ and Higgsinos.
These mixings bring the suitable contributions 
to $\bar N_\alpha$ Majorana mass.
This requires $\lambda_9^{331}\sim 10^{-5}$ and 
$\lambda_9^{332} \sim 10^{-4}$.\footnote{This estimation also 
depends on a value of 
$\langle H_3^1\rangle$. However, in this model there are two Higgs
doublet pairs and then it should take a rather small VEV.}
For these parameters the conditions (21)$\sim$(23) seems to be 
generally satisfied.  
As easily seen, however, the estimations of eqs.(20) are dependent on
the assumptions of the soft supersymmetry breaking parameters. 
If we loose these assumptions and consider more general situations, 
their numerical estimations can rather largely change and the reqiured 
conditions may also be changed to realize the suitable hierarchy. 
In any case we may have to consider the effects of some nonuniversal
structure of soft supersymmetry breaking
parameters\cite{nonuni} and also the multi-Higgses to know whether 
we can get the favorable mass 
hierarchy in the neutrino sector of this model.
It will be necessary to practice more quantitative study of this point
providing the structure of soft supersymmetry breaking parameters.

\section{Discussions and summary}
We have studied that our present model works well in the 
triplet-doublet splitting
and the small neutrino mass generation in addition that it has the similar
structure to the MSSM.
Now we order some other brief comments which should be added on its
phenomenological features.\\
(1)~It is well known that the existence of the light singlet fields is 
dangerous because their couplings to heavy fields can induce the vaccum 
instability through the tad pole diagram\cite{tad}. However, in our model the
light singlets $S_2$ and $S_3$ can not couple to heavy fields and then
this model is free from the tadpole problem.
This decoupling of $S_2$ and $S_3$ from heavy fields are due to the 
discrete symmetry and the unconventional field assignment.\\
(2)~In the string inspired $E_6$ models, there is no $E_6$ relation
among Yukawa couplings\cite{w}.
This fact makes Yukawa couplings of extra color triplets $g$ and $\bar g$
with ordinary matters and the singlet ${\cal J}$ to have less constraints than
those in the usual supersymmetric GUTs.
Because of the ambiguity caused by this looseness we can not definitely 
estimate the lower bound of the VEV 
$\langle{\cal J}\rangle$ from the experimental bound of
proton decay as done in refs.\cite{proton}. 
However, we can always set up $\langle{\cal J}\rangle$ to escape
its bound without bringing any other problems.\\ 
(3)~The study of radiative symmetry breaking at $\sim$1~TeV scale 
based on the renormalization group equations is an
interesting problem.
In such a study experimental bounds including the top mass give us
various informations on the parameters in the model, especially,
on the extra $Z$ mass as found in \cite{sy2}.
As suggested in \cite{nar}, there is the nonuniversal neutral current
interaction due to the extra $Z$ in the present model. 
However, the extra $Z$ mass is expected to be ${^>_\sim}~1$~TeV from 
the consistency with the primodial
nucleosynthesis\cite{wssok} because of the existence of extra light neutrino
species. Although its observation will not be expected in near future,
the consistency check of such extra $Z$ mass with radiative symmetry breaking
is interesting.\\
(4)~To make a realistic model on this line there remain many problems to
be considered, for example, the derivation of the realistic 
quark/lepton mass matrices.
It will be necessary again to impose suitable discrete symmetries which act
nontrivially on the quark sector to overcome this problem.
One promising possibility is to give the Yukawa couplings in the quark sector
in the following way,
\begin{equation}
\lambda_2^{ij2}\left({{\cal J}\bar{\cal J} \over M_{\rm pl}^2}
\right)^{n^U_{ij}}Q_i\bar U_jH_2^2, \qquad
\lambda_4^{ij2}\left({{\cal J}\bar{\cal J} \over M_{\rm pl}^2}
\right)^{n^D_{ij}}Q_i\bar D_jH_2^1, 
\end{equation}
where $n^U_{ij}$ and $n^D_{ij}$ are zero or positive integers and their
values are determined by the discrete symmetries.
This scheme is very similar to the proposal in \cite{ir}.
However, the extra $U(1)$ invariance requires the appearence of 
${\cal J}\bar{\cal J}$ pair in these formulae so that $\langle {\cal J}\rangle$
should be rather large value $|\langle {\cal J}\rangle|\sim 0.2M_{\rm pl}$
to realize the correct hierarchy under the assumption for the Yukawa couplings
$\lambda \sim O(1)$.\\
(5)~The right handed neutrino mass production through an abelian 
D-flat direction may 
be related to the inflation and also the primodial baryon number asymmetry
as suggested in \cite{sy1}. The study of this aspect is also necessary.  
 
In summary we studied the unconventional field assignment in
string inspired $E_6$ models with a conjugate pair of chiral superfields
$({\cal J}, \bar{\cal J})$
which are $G_{\rm SM}$ singlets. Extra color triplets become heavy
enough to guarantee the proton stability through VEVs of these singets,
although doublet Higgs fields are kept light.
The massless fields sector can be almost the MSSM by imposing suitable
discrete symmetries, except for neutralino, chargino and neutrino sectors.
We showed that the interesting neutrino mass matrix 
can be derived in these models.
These mass matrices may be able to explain the hierarchical masses
appropriate for solar neutrino, atmospheric neutrino and dark matter.
We consider here the restricted discrete symmetries which satisfy certain 
conditions.
If we loose these conditions, there will be many other possibilities which may
present fruitful neutrino mass structures and also other interesting 
phenomenology.
Anyway, the possibility proposed here to prohibit the fast proton 
decay and give neutrinos small masses simultaneously 
seems to be worthy for further study.\\

\noindent
%{\Large\bf Acknowledgement}\\
The author is grateful for the hospitality of TH-Division of CERN
where the part of this work was done.
This work is supported in part by a Grant-in-Aid for Scientific 
Research from the Ministry of Education, Science and Culture
(\#05640337).

\newpage
%%%%%%%%%%%%%%%%%%%%%%%%%%%% References %%%%%%%%%%%%%%%%%%%%%%%%%%%%%%%%%%%

\newpage

{\Large\bf Table 1}\\
The decomposition of ${\bf 27}$ of $E_6$ under $G_{\rm SM}$
and the field assignment for them. All $U(1)$ charges are normalized
as $\displaystyle \sum_{i \in {\bf 27}}Q_i^2=5$.
\vspace{15mm}
\begin{center}
\begin{tabular}{|c||c|c|c|c|c|}\hline
fields &$G_{\rm SM}$& ${12 \over \sqrt{10}}Q_\psi$ & 
${12 \over \sqrt 6}Q_\chi$ & $i=1,2$ &$i=3$ \\ \hline\hline
$A_i$ &$(3,2)_{1 \over 6}$& 1 & $-1$ &$Q$& $Q$ \\
$B_i$ &$(3^*,1)_{-2 \over 3}$& 1 & $-1$ &$\bar U$& $\bar U$ \\
$C_i$ &$(3^*,1)_{1 \over 3}$ & $-2$ &$-2$ &$\bar g$& $\bar g$ \\
$D_i$ &$(3^*,1)_{1 \over 3}$ & 1 & 3 & $\bar D$ & $\bar D$ \\
$E_i$ &$(3,1)_{-1 \over 3}$& $-2$ & 2 & $g$ & $g$ \\
$F_i$ &$(1,2)_{1 \over 2}$&  $-2$ & 2 & $H^2$ & $H^2$ \\
$G_i$ &$(1,2)_{-1 \over 2}$& 1 & 3 & $L$ & $H^1$ \\
$H_i$ &$(1,2)_{-1 \over 2}$ & $-2$ &$-2$ &$H^1$& $L$ \\
$I_i$ &$(1,1)_1$& 1 &$-1$ & $\bar E$ & $\bar E$ \\
$J_i$ &$(1,1)_0$& 4 &0 & $S$ & $\bar N$ \\
$K_i$ &$(1,1)_0$& 1& $-5$& $\bar N$ & $S$ \\ \hline
\end{tabular}
\end{center}
\newpage
{\Large\bf Figure Captions}\\

\noindent
{\Large\bf Fig. 1}\\
Supergraph of one-loop $L_\alpha \bar N_3$ Dirac neutrino mass.
Either vertex of $\lambda_{10}^{(23)22}$ or $\lambda_{11}^{33i}$ should 
be understood as a soft supersymmetry breaking A-term.\\

{\Large\bf Fig. 2}\\
Supergraph of one-loop $\bar N_\alpha \bar N_3$ Majorana neutrino mass.
Either vertex of $\lambda_9^{(23)22}$ or $\lambda_9^{(23)33}$ should be 
understood as a soft supersymmetry breaking A-term.\\

{\Large\bf Figure 3}\\
Supergraph of one-loop $\bar N_\alpha \bar N_\beta$ Majorana neutrino mass.
Either vertex of $\lambda_9^{(23)33}$ should be understood as 
a soft supersymmetry breaking A-term.

\newpage
\vspace*{1.5cm}
\begin{center}
\begin{picture}(230,160)
\setlength{\unitlength}{1mm}
\thicklines
\put(40,40){\oval(40,40)[lb]}
\put(40,40){\oval(40,40)[lt]}
\put(40,40){\oval(40,40)[rb]}
\put(40,40){\oval(40,40)[rt]}
\put(0,40){\line(10,0){20}}
\put(60,40){\line(10,0){20}}
\put(40,10){\line(0,10){10}}
\put(40,60){\line(0,10){10}}
\put(38,8){$\bigotimes$}
\put(38,70){$\bigotimes$}
\put(1,32){\Large\bf  $L_\alpha$}
\put(75,32){\Large\bf $\bar N_3$}
\put(35,1){\Large\bf $\langle H_3^1\rangle$}
\put(35,78){\Large\bf $\langle S_2\rangle$}
\put(21,40){\large\bf $\lambda_{11}^{\alpha 2i}$}
\put(35,53){\large\bf $\lambda_{10}^{(23)22}$}
\put(36,22){\large\bf $\lambda_{11}^{33i}$}
\put(47,40){\large\bf $\lambda_{10}^{(23)33}$}
\put(18,62){\Large\bf $H_2^1$}
\put(13,20){\Large\bf $\bar E_i$}
\put(60,60){\Large\bf $\left(\begin{array}{cc}H_2^2\\ H_3^2 
\end{array}\right)$}
\put(60,20){\Large\bf $L_3$}
\end{picture}\\
\vspace{.6cm}
{\Large\bf Fig.1}\vspace{4cm}
\end{center}
\begin{center}
\begin{picture}(230,160)
\setlength{\unitlength}{1mm}
\thicklines
\put(40,40){\oval(40,40)[lb]}
\put(40,40){\oval(40,40)[lt]}
\put(40,40){\oval(40,40)[rb]}
\put(40,40){\oval(40,40)[rt]}
\put(0,40){\line(10,0){20}}
\put(60,40){\line(10,0){20}}
\put(40,10){\line(0,10){10}}
\put(40,60){\line(0,10){10}}
\put(38,8){$\bigotimes$}
\put(38,70){$\bigotimes$}
\put(1,32){\Large\bf  $\bar N_\alpha$}
\put(75,32){\Large\bf $\bar N_3$}
\put(35,1){\Large\bf $\langle S_2\rangle$}
\put(35,78){\Large\bf $\langle S_3\rangle$}
\put(21,40){\large\bf $\lambda_{9}^{(23)3\alpha}$}
\put(35,53){\large\bf $\lambda_{10}^{(23)33}$}
\put(36,22){\large\bf $\lambda_{10}^{(23)22}$}
\put(47,40){\large\bf $\lambda_{10}^{(23)23}$}
\put(18,62){\Large\bf $H_3^1$}
\put(3,13){\Large\bf $\left(\begin{array}{cc}H_2^2\\ H_3^2 
\end{array}\right)$}
\put(60,60){\Large\bf $\left(\begin{array}{cc}H_2^2\\ H_3^2 
\end{array}\right)$}
\put(60,20){\Large\bf $H^1_2$}
\end{picture}\\
\vspace{.6cm}
{\Large\bf Fig.2}
\end{center}
\newpage
\vspace*{1.5cm}
\begin{center}
\begin{picture}(230,160)
\setlength{\unitlength}{1mm}
\thicklines
\put(40,40){\oval(40,40)[lb]}
\put(40,40){\oval(40,40)[lt]}
\put(40,40){\oval(40,40)[rb]}
\put(40,40){\oval(40,40)[rt]}
\put(0,40){\line(10,0){20}}
\put(60,40){\line(10,0){20}}
\put(40,10){\line(0,10){10}}
\put(40,60){\line(0,10){10}}
\put(38,8){$\bigotimes$}
\put(38,70){$\bigotimes$}
\put(1,32){\Large\bf  $\bar N_\alpha$}
\put(75,32){\Large\bf $\bar N_\beta$}
\put(35,1){\Large\bf $\langle S_3\rangle$}
\put(35,78){\Large\bf $\langle S_3\rangle$}
\put(21,40){\large\bf $\lambda_{9}^{(23)3\alpha}$}
\put(35,53){\large\bf $\lambda_{9}^{(23)33}$}
\put(36,22){\large\bf $\lambda_{9}^{(23)33}$}
\put(46,40){\large\bf $\lambda_{9}^{(23)3\beta}$}
\put(18,62){\Large\bf $H_3^1$}
\put(3,13){\Large\bf $\left(\begin{array}{cc}H_2^2\\ H_3^2 
\end{array}\right)$}
\put(60,60){\Large\bf $\left(\begin{array}{cc}H_2^2\\ H_3^2 
\end{array}\right)$}
\put(60,20){\Large\bf $H^1_3$}
\end{picture}\\
\vspace{.6cm}
{\Large\bf Fig.3}
\end{center}
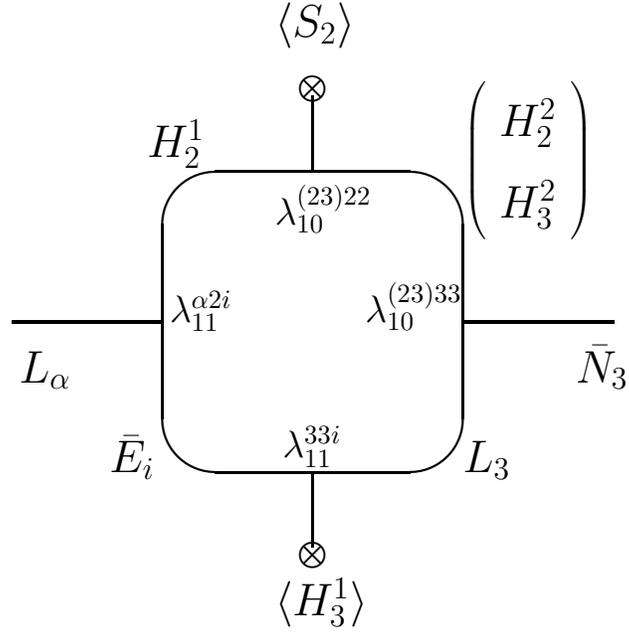
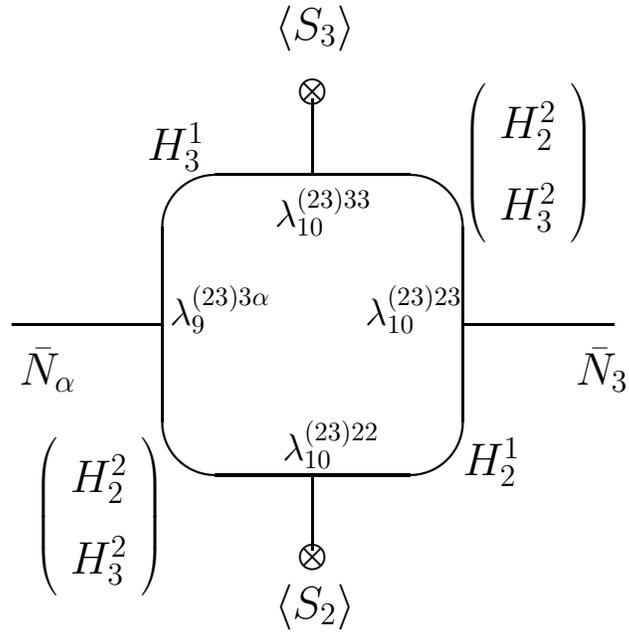
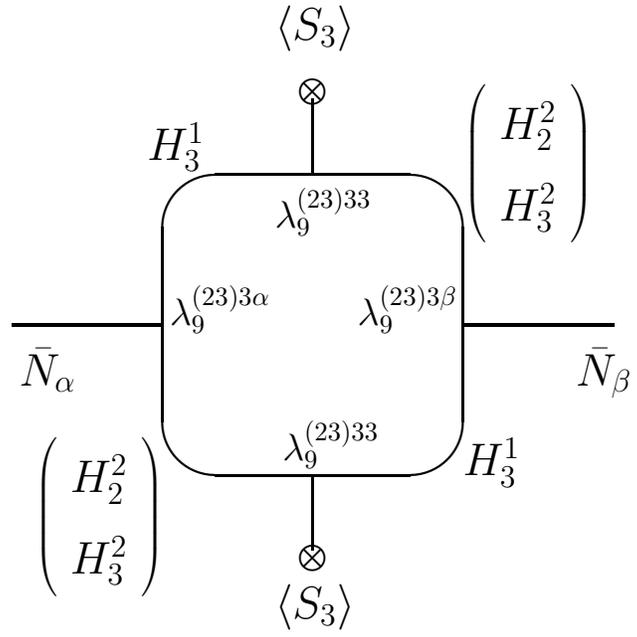
\end{document}